\documentclass[aps,pra,reprint,superscriptaddress,floatfix]{revtex4-2}

\usepackage{xcolor}
\usepackage{amsmath}
\usepackage{amssymb}
\usepackage{amsfonts}
\usepackage{graphicx}
\usepackage{xspace} 
\usepackage{float}
\usepackage{siunitx} 
\usepackage[version=4]{mhchem}
\usepackage{soul}
\usepackage{esint}
\usepackage{bm}
\usepackage{xr-hyper}
\usepackage{textcomp}

\usepackage[mathlines]{lineno}
\usepackage{enumerate}
\usepackage[unicode=true]{hyperref}
\hypersetup{colorlinks=true, linkcolor=blue, citecolor=blue, urlcolor=blue}

\usepackage[normalem]{ulem}

\usepackage{silence}
\begin{document}
\title{Anisotropic interface-confined superconductivity in \ce{FeTe}-based heterostructures}
\author{Bicky Singh Moirangthem}
\affiliation{Ames National Laboratory, Ames, IA 50011, U.S.A.}
\affiliation{Department of Physics and Astronomy, Iowa State University, Ames, IA 50011, U.S.A.}
\author{Kamal R. Joshi}
\affiliation{Ames National Laboratory, Ames, IA 50011, U.S.A.}
\author{Zi-Jie Yan}
\affiliation{Department of Physics, The Pennsylvania State University, University Park, PA 16802, U.S.A.}
\author{Pu Xiao}
\affiliation{Department of Physics, The Pennsylvania State University, University Park, PA 16802, U.S.A.}
\author{Lok-Kan Lai}
\affiliation{Department of Physics, The Pennsylvania State University, University Park, PA 16802, U.S.A.}
\author{Makariy A. Tanatar}
\affiliation{Ames National Laboratory, Ames, IA 50011, U.S.A.}
\affiliation{Department of Physics and Astronomy, Iowa State University, Ames, IA 50011, U.S.A.}
\author{Cui-Zu Chang}
\affiliation{Department of Physics, The Pennsylvania State University, University Park, PA 16802, U.S.A.}
\author{Ruslan Prozorov}
\email{Corresponding author: prozorov@ameslab.gov}
\affiliation{Ames National Laboratory, Ames, IA 50011, U.S.A.}
\affiliation{Department of Physics and Astronomy, Iowa State University, Ames, IA 50011, U.S.A.}

\date{29 September 2026}

\begin{abstract}
Interface-confined superconductivity emerges from the interaction of electronic states across chemically distinct boundaries, providing a route to engineer superconducting phases where magnetism and topology coexist. Determining the intrinsic nature of such superconductivity, however, is challenging because the superconducting layer is only a few nanometers thick and buried beneath several normal layers. Here, we measure the whole-sample Meissner response of six $\mathrm{FeTe}$-based heterostructures in a uniform magnetic field using a frequency-domain tunnel-diode resonator. In the ultrathin limit, the conventional normalization of the measured susceptibility, $\chi(T\to0)=-1$, fails by tens of percent. We establish the appropriate calibration and invert $\chi(T)$ to determine the London penetration depth $\lambda(T)$. Two key results emerge. First, the broad transitions observed in $\chi(T)$ arise naturally from the extreme geometry and large $\lambda$, without requiring chemical or structural inhomogeneity; the extracted $\lambda(T)$ closely tracks the resistive transition. Second, $\lambda(T)$ and the corresponding superfluid density are inconsistent with a fully gapped isotropic $s$-wave state and instead indicate a strongly anisotropic order parameter possibly with line nodes or deep gap minima. The inferred $\lambda(0)$ is of order $1\,\mu$m, consistent with an independent analysis of the 2D phase stiffness. Despite the distinct chemical, magnetic, and topological character of the three overlayers, all six FeTe heterostructures exhibit similar low-temperature power-law behavior, with no systematic dependence of the superconducting response on overlayer identity. These results point to the interfacial $\mathrm{FeTe}$ layer as the common origin of superconductivity.
\end{abstract}

\maketitle

\section{Introduction}
Interfacial superconductivity has emerged as an important route to creating superconducting states that are absent in the constituent materials. Unlike conventional superconductivity in bulk crystals and homogeneous thin films, interfacial superconductivity can be confined to an ultrathin 2D region at the boundary between chemically distinct materials. Remarkably, superconductivity appears in systems in which none of the constituents is superconducting on its own: the $\mathrm{LaAlO}_3/\mathrm{SrTiO}_3$ interface between two band insulators \cite{ohtomo2004,reyren2007} and the $\mathrm{EuO}/\mathrm{KTaO}_3$ and $\mathrm{LaAlO}_3/\mathrm{KTaO}_3$ interfaces \cite{liu2021}. It also appears in artificially layered metallic films, such as $\mathrm{Au/Ge}$ multilayers \cite{seguchi1990}, and, with a strongly enhanced transition temperature, in $\mathrm{La}_{1.55}\mathrm{Sr}_{0.45}\mathrm{CuO}_4/\mathrm{La}_2\mathrm{CuO}_4$ bilayers \cite{gozar2008} and in monolayer $\mathrm{FeSe}/\mathrm{SrTiO}_3$ \cite{wang2012,zhang2014,ge2015}.

Furthermore, the superconducting transition temperature, $T_c$, can be significantly enhanced compared with its bulk value. For example, bulk $\mathrm{FeSe}$ has superconducting $T_c \approx 8~\mathrm{K}$ \cite{hsu2008}, whereas a single unit cell of $\mathrm{FeSe}$ grown on $\mathrm{SrTiO}_3$ shows $T_c \approx 40~\mathrm{K}$ in transport \cite{zhang2014} and a spectroscopic gap surviving above $50~\mathrm{K}$ \cite{wang2012}.
These observations demonstrate the potential of interface engineering to enhance superconductivity and stabilize emergent quantum states. Exploiting this potential, however, requires identifying the microscopic mechanisms responsible for interfacial superconductivity, including changes in chemical composition, strain, charge transfer, spin texture, and interfacial phonons \cite{wang2016,liu2024,maggiora2024}. One of the prominent emergent phenomena is topological superconductivity. Beyond testing an otherwise largely abstract theory, a material realization is sought for its potential in fault-tolerant qubits based on Majorana bound states \cite{alicea2012,sato2017,nayak2008}.
One proposed approach to realize this is to bring a superconductor into proximity with a topological insulator (TI) \cite{fu2008,qi2011}. Over the past decade, interfacial superconductivity has been observed in several TI/$\mathrm{FeTe}$ heterostructures, including (Bi,Sb)$_2$Te$_3$/FeTe, MnBi$_2$Te$_4$/FeTe, Cr-doped (Bi,Sb)$_2$Te$_3$/FeTe, and (Pb,Sn)Te/FeTe \cite{he2014,yuan2024coexistence,liang2020,manna2017,qin2020,moore2023,yi2024,yi2023dirac,yan2026interface}, providing a platform for the interplay between superconductivity, magnetism, and topology. The microscopic origin of superconductivity in these FeTe-based heterostructures, however, remains unresolved. A study of $1\mathrm{T}-\mathrm{CrTe}_2/\mathrm{FeTe}$ heterostructures reveals that superconductivity persists even when the top layer is topologically trivial \cite{yan2025meissner}. Another study suggests that Te stoichiometry at the interface plays a crucial role in the emergence of this phenomenon \cite{yao2025mystery}.

Recently, Yan \textit{et al.} \cite{yan2026stoichiometric} demonstrated that Te-flux annealing removes interstitial Fe from molecular-beam-epitaxy (MBE) grown films, producing stoichiometric $\mathrm{FeTe}$ with $T_c \approx 10-13.5~\mathrm{K}$ even in the absence of an overlayer. This temperature range closely match the $T_c \approx 9.5-12~\mathrm{K}$ observed in the FeTe-based heterostructures studied here.
Very recently, a combined scanning-superconducting quantum interference device (SQUID), scanning tunnelling microscopy and spectroscopy (STM/S), and transport study of stoichiometric FeTe reported a non-saturating London penetration depth down to $0.02\,T_c$ with a power-law exponent $\approx 1-1.5$ and a V-shaped low-energy density of states, concluding a gap with nodes or deep minima \cite{li2026nodal}.
Despite these studies, the whole-sample magnetic response and mechanisms of interfacial superconductivity in $\mathrm{FeTe}-$based heterostructures remain experimentally unexplored.

Most studies of interfacial superconductivity employ microscopic (e.g., STM/STS \cite{wang2012,moore2023,yuan2024coexistence,dai2017}) and mesoscopic (e.g., SQUID, Magnetic Force Microscopy \cite{bert2011,yan2025meissner,bert2012,li2026nodal}) local probes. However, to date, there have been only limited reports on the whole-sample response of the superconducting state \cite{singh2018,mallik2022,gasparov2012}. It is challenging to probe the magnetic response of an ultrathin film due to competing factors --- for a thin platelet in a perpendicular field, the magnetic moment scales as the cube of the lateral size, $m \propto R^3$, and is independent of the thickness in the ideal-screening limit, so the lateral dimensions should be maximized --- but then the sample no longer fits most magnetometers.
The closest technique is two-coil mutual inductance \cite{turneaure1996,turneaure1998,lemberger2007,zhang2021}, but it is not truly global: the drive coil is necessarily small compared with the film to avoid sample edges, so the sample is probed by a strongly inhomogeneous field and the measured mutual inductance is a coil-weighted local average rather than a whole-sample susceptibility. The required drive amplitude is relatively large since the measurements are done in the amplitude domain.

We have developed a frequency-domain susceptometer based on a tunnel-diode resonator \cite{Van1975,prozorov2000a,prozorov2011,prozorov2021,giannetta2022london}, designed specifically for thin films, $f$TDR.
It has been successfully benchmarked in a study of conventional tantalum thin films \cite{moirangthem2026}. The device measures the total magnetic susceptibility of the entire film and is particularly useful for probing the magnetic response of interfacial layers buried between two different materials, which makes using surface probes challenging.
To the best of our knowledge, the present work is the first report of measuring interfacial superconductivity through the whole-sample London-Meissner response in an applied weak magnetic field — the linear-response limit.
Moreover, we describe the novel calibration protocol to extract the London penetration depth, $\lambda(T)$, from the measured whole-sample magnetic susceptibility, $\chi(T)$. In six $\mathrm{FeTe}$-based heterostructures, we observed robust Meissner screening and concluded that the superconducting order parameter is highly anisotropic, likely nodal.
\begin{figure*}[tb]
\includegraphics[width=17cm]{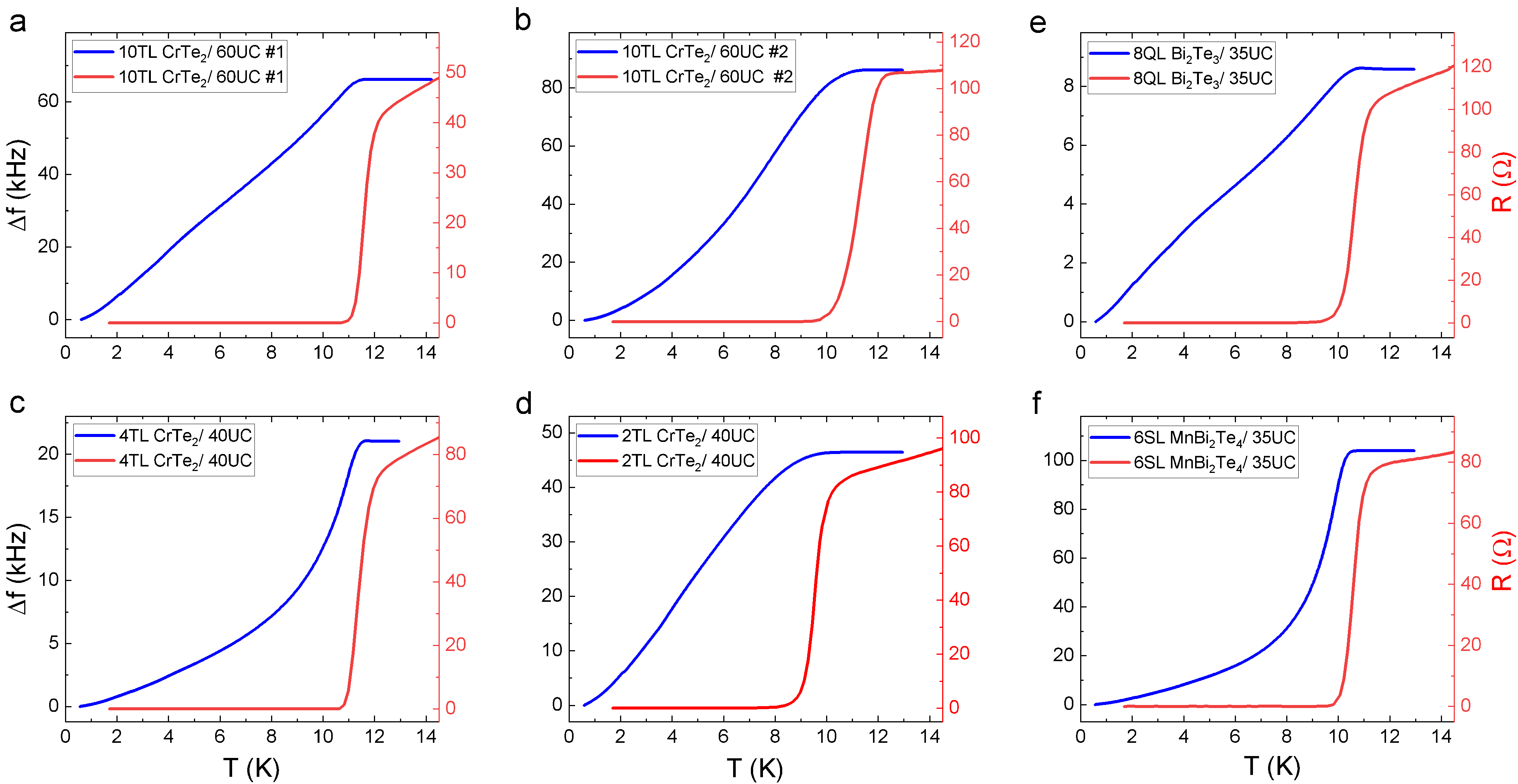}
\caption{(a)-(f) Comparison of temperature dependent resonant frequency change $\Delta f$ and measured resistance $R$ in six $\mathrm{FeTe}$ heterostructures. The onsets of superconducting transition observed in transport measurements are fairly consistent with the appearance of diamagnetic screening in all six samples.}
\label{fig:df(T)_and_R(T)}
\end{figure*}

\section{Studied Materials}
Conventional as grown $\mathrm{FeTe}$ containing interstitial Fe is a non-superconducting antiferromagnet, with $\mathrm{Fe}$ moments arranged in a bi-collinear magnetic structure in the $ab$-plane. It undergoes coupled magnetic and structural transitions near $50-70$ K, where the crystal symmetry changes from tetragonal to monoclinic (for small interstitial Fe content; larger excess Fe drives the low-temperature structure to be orthorhombic and the magnetic order incommensurate) \cite{enayat2014,bao2009,li2009}. Despite the absence of superconductivity in bulk $\mathrm{FeTe}$, superconductivity can emerge when $\mathrm{FeTe}$ films are interfaced with a variety of van der Waals materials.

In this work, we investigate three classes of $\mathrm{FeTe}$-based heterostructures with distinct overayers: $1\mathrm{T}-\mathrm{CrTe}_2/\mathrm{FeTe}$, $\mathrm{MnBi}_2\mathrm{Te}_4/\mathrm{FeTe}$, and $\mathrm{Bi}_2\mathrm{Te}_3/\mathrm{FeTe}$, previously characterized in Refs.~\cite{yan2025meissner,yuan2024coexistence,yi2023dirac}. These materials provide three distinct combinations of magnetism and topology: $\mathrm{Bi}_2\mathrm{Te}_3$ is a nonmagnetic topological insulator \cite{chen2009}, $\mathrm{MnBi}_2\mathrm{Te}_4$ is both magnetic and topological \cite{Li2019,Zhang2019,Otrokov2019}, whereas $1\mathrm{T}-\mathrm{CrTe}_2$ is ferromagnetic without known nontrivial band topology \cite{freitas2015}. 
The heterostructures consist of $\mathrm{FeTe}$ interfaced with several quintuple layers (QLs) of $\mathrm{Bi}_2\mathrm{Te}_3$, septuple layers (SLs) of $\mathrm{MnBi}_2\mathrm{Te}_4$, or trilayers (TLs) of $\mathrm{CrTe}_2$. All samples were grown by MBE on heat-treated SrTiO$_3$ (100) substrates and exhibit atomically sharp interfaces. More details on the MBE growth and structural characterization are provided in Refs.~\cite{yan2025meissner,yuan2024coexistence,yi2023dirac}.

\begin{figure}[tb]
\includegraphics[width=8cm]{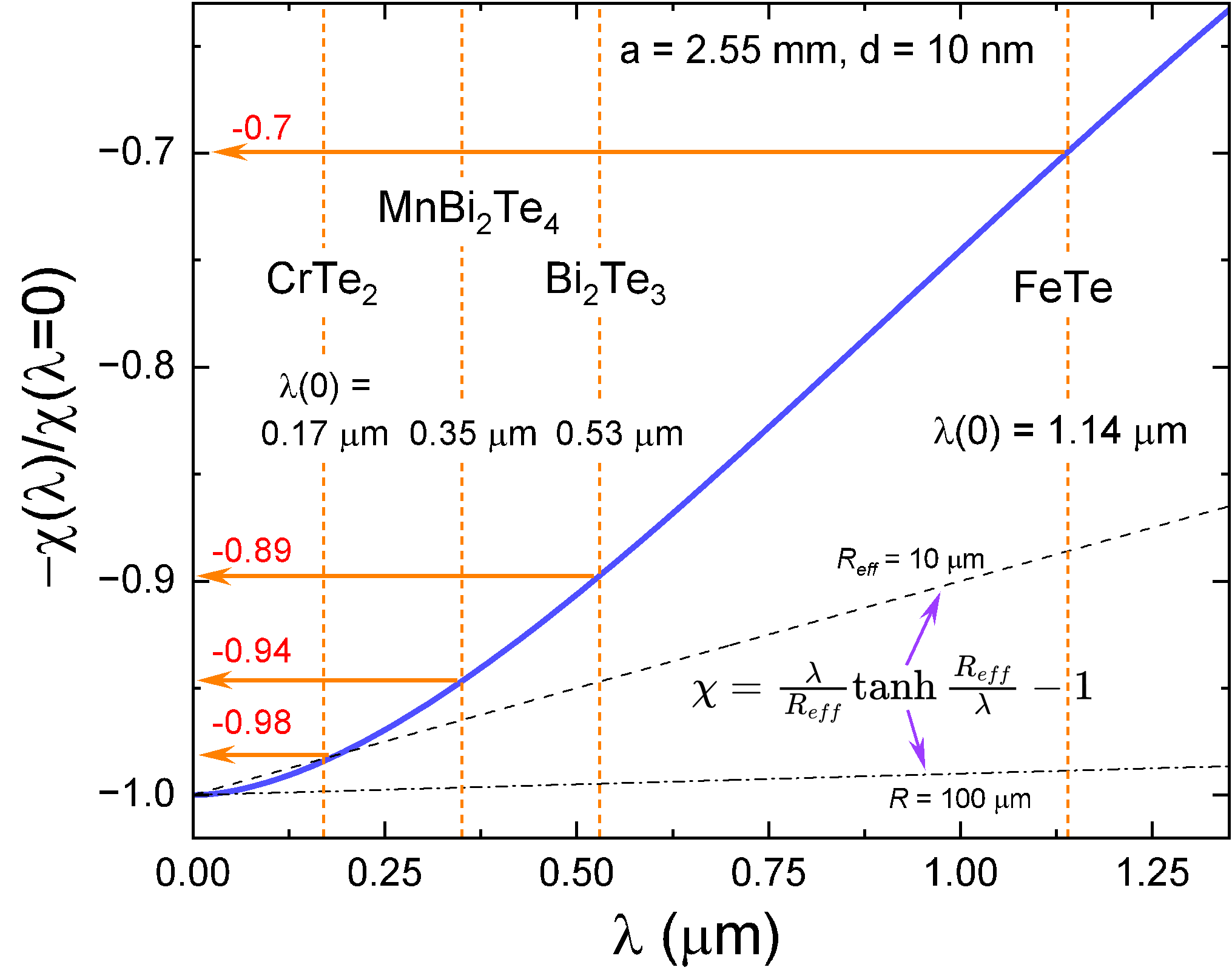}
\caption{The normalized magnetic susceptibility as a function of the London penetration depth $\chi(\lambda)$ calculated using COMSOL finite element software for our sample dimensions in a perpendicular applied magnetic field. For comparison, dashed and dashed-dotted lines show ``conventional'' magnetic susceptibility used to describe usual, even thin, crystals with the effective dimension, $R_{eff}$, absorbing the demagnetizing factor corrections \cite{prozorov2000,prozorov2021}. 
For thicker crystals, typical $R_{eff} \sim 100\,\mu\mathrm{m}$ and for very thin samples, $R_{eff} \sim 10\,\mu\mathrm{m}$. In either case, there is an obvious and significant difference with the ultrathin limit $\chi(\lambda)$. The arrows and numbers show how to obtain the normalization values from known $\lambda(0)$, estimated for our compounds in Table~\ref{tab:lambda}. \textit{Knowing this calibration value is key to proper extraction of $\lambda(T)$ from $\chi(T)$ in ultrathin limit.}
}
\label{fig:chi(lambda)}
\end{figure}
\section{Global magnetic susceptibility in the ultrathin limit}
\label{sec:calib}

Six $\mathrm{FeTe}$ heterostructures, four with $1\mathrm{T}-\mathrm{CrTe}_2$ overlayers and one each with $\mathrm{Bi}_2\mathrm{Te}_3$ and $\mathrm{MnBi}_2\mathrm{Te}_4$ overlayers, were measured in a $f$TDR down to 500 mK. 
Figure \ref{fig:df(T)_and_R(T)} compares the resonance-frequency shift $\Delta f$ with the corresponding resistance $R (T)$. In all FeTe-based heterostrutures, $\Delta f(T)$ saturates upon entering the normal state, with the onset of diamagnetic screening closely matching the onset of the resistive transition. In contrast to the sharp drop of $R (T)$ below $T_c$, however, $\Delta f(T)$, which is proportional to the dynamic magnetic susceptibility, evolves gradually over a broad temperature range below $T_c$. Because superconductivity is confined to an ultrathin interfacial layer only a few nanometers thick, such a broad magnetic transition might conventionally be attributed to structural or chemical inhomogeneity, granularity, or disorder. We show instead that this behavior arises naturally in the ultrathin limit, even for an intrinsically homogeneous superconductor.
\begin{table}[h]
\centering
\begin{tabular}{lccc}
\hline
Material &
$m^*/m_e$ &
$n_e$ (cm$^{-3}$) & $\lambda(0)$ \\
\hline
FeTe & 46 \cite{lin2025,tamai2010} & $10^{21}$ \cite{tsukada2011} & $1.14\,\mu\mathrm{m}$ \\
CrTe$_2$ & 0.99 \cite{rasmussen2015} & $10^{21}$ \cite{meng2021} & $170\,\mathrm{nm}$ \\
Bi$_2$Te$_3$ & 0.10 \cite{kamenskyi2020} & $10^{19}$ \cite{marchenkov2023} & $0.53\,\mu\mathrm{m}$ \\
MnBi$_2$Te$_4$ & 0.043 \cite{lei2022} & $10^{19}$ \cite{hu2021} & $0.35\,\mu\mathrm{m}$ \\
\hline
\end{tabular}
\caption{London penetration depths estimated from the Drude--London theory: $\lambda=\sqrt{m^\star/n_e\mu_0 e^2}$ for $\mathrm{FeTe}$, CrTe$_2$, Bi$_2$Te$_3$, and MnBi$_2$Te$_4$ using carrier concentration from references shown.
None of the three overlayer materials is superconducting; these values are quoted only to bracket a physically plausible range of carrier concentrations, hence $\lambda(0)$ for the interfacial condensate.}
\label{tab:lambda}
\end{table}
In samples with demagnetizing factor $N$, the measured signal --- voltage in the case of amplitude domain susceptometers or the resonant frequency shift in microwave cavity perturbation technique (and used here tunnel-diode resonator), is enhanced by a factor of $1/(1-N)$. To infer the intrinsic behavior, traditionally, such data are normalized between -1 and 0 to convert it into magnetic susceptibility, assuming that the sample becomes a perfect diamagnet at $T \to 0$. Then, to extract the London penetration depth, the ``effective'' sample dimension, $R_{eff}$, is calculated from the sample geometry \cite{prozorov2000,prozorov2021,prozorov2023}. This approach has been successfully used to study crystals, even quite thin, down to tens of micrometers \cite{prozorov2011,prozorov2021}. More precisely, $(1-N)\chi(T \to 0 ) \approx -1+\lambda(0)/R_{eff}$, but in most cases $\lambda(0) \ll R_{eff}$, with $R_{eff}$ in the range, $10 - 500\,\mu\mathrm{m}$, and this small deviation does not affect the resulting temperature dependence of the extracted superfluid stiffness, which is then used to analyze the superconducting gap structure.
\begin{figure}[tb]
\includegraphics[width=8cm]{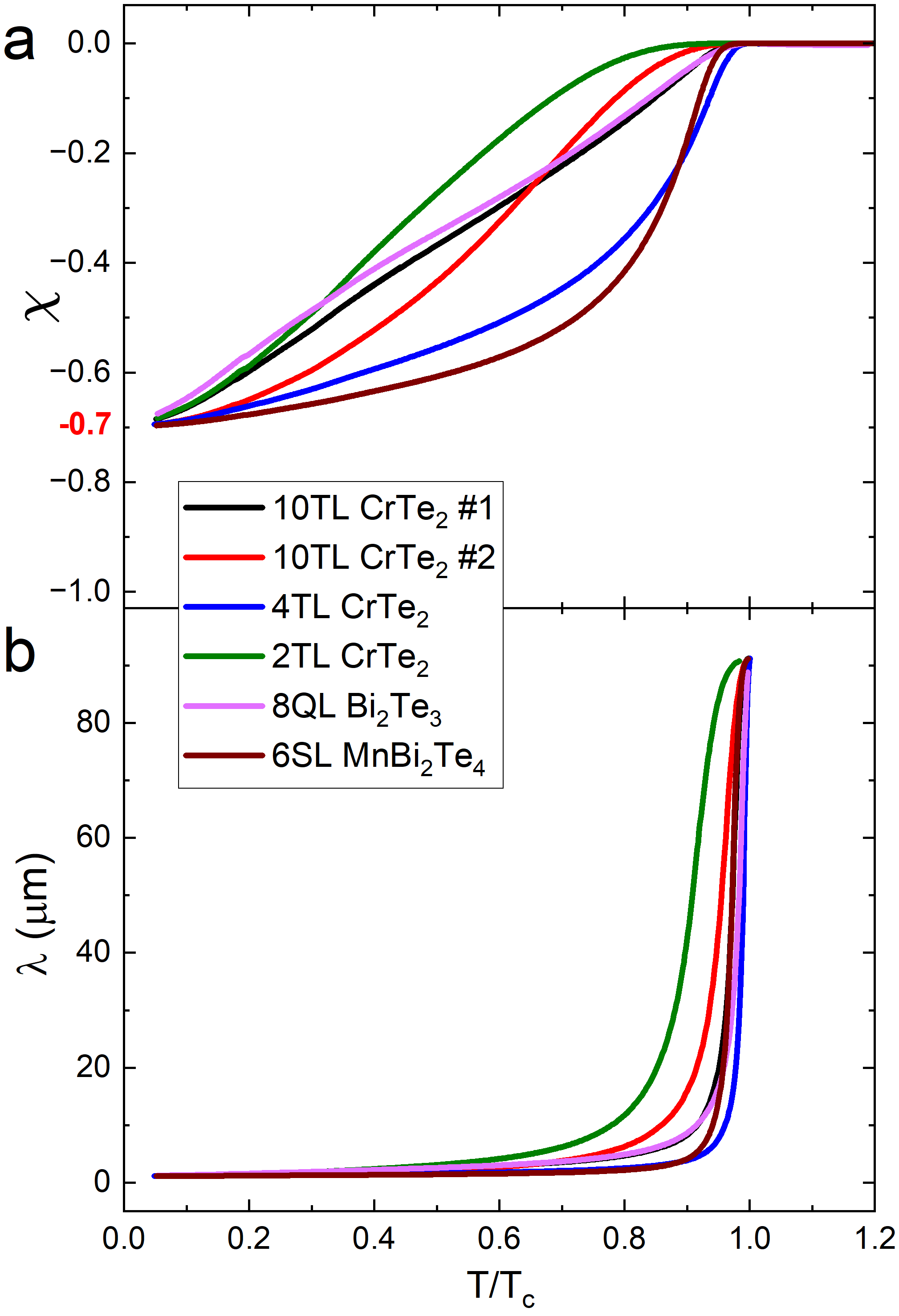}
\caption{(a) The variation of normalized magnetic susceptibility $\chi(T/T_c)$ in six measured $\mathrm{FeTe}$ heterostructures. Considering the $\lambda(T=0)\approx$ $1.14\,\mu\mathrm{m}$ for $\mathrm{FeTe}$, $\chi$ has been normalized between -0.7 to 0. (b) Corresponding London penetration depth $\lambda$ extracted using the calibration curve in Fig.\ref{fig:chi(lambda)}.}
\label{fig:chi(t)_and_lambda(t)}
\end{figure}
The situation in an ultrathin film in a \emph{perpendicular} field is different, and The origin of this difference is the relevant screening length. Screening of a film of thickness $d\ll\lambda$ is governed not by $\lambda$ but by the Pearl screening length $\Lambda=\lambda^{2}/d$~\cite{pearl1964}, which for our films exceeds $\lambda$ by three orders of magnitude. The parallel-field case is qualitatively different and far less severe: the exact slab result $\chi=(2\lambda/d)\tanh(d/2\lambda)-1$ reduces to $\chi\simeq-d^{2}/12\lambda^{2}$ for $d\ll\lambda$~\cite{krieger1964}. 
The corresponding integral equation for the sheet current was solved by matched asymptotics for discs and strips by Dorsey~\cite{dorsey1995}, who recovered the ideal-screening solution $K(r)\propto r/\sqrt{a^{2}-r^{2}}$~\cite{mikheenko1993,clem1994}, and Brandt gave a general numerical scheme valid for arbitrary film shape and arbitrary $\Lambda$~\cite{brandt2005}. The essential point for what follows is that the approach to perfect screening is controlled by $\Lambda/a$ and not by $\lambda/R$, and is therefore slower by a factor of order $\lambda/d\sim10^{2}$.

The magnitude of this geometric effect can be illustrated quantitatively. We model the measured film as a disc of thickness $d=10$~nm, because the reported thickness of the interfacial superconducting layer in FeTe heterostructures lies between $6$ and $10$~nm~\cite{he2014,yi2023dirac,yan2025meissner,yuan2024coexistence,yi2024,balakrishnan2025}. The ideal-screening susceptibility of this disc is $\chi(\lambda\to0)=-8a/3\pi d=-2.16\times10^{5}$, corresponding to a demagnetizing factor $N=1-4.6\times10^{-6}$. Even at this extreme, a perfect diamagnet still satisfies $(1-N)\chi(T\to0,\lambda\to0)=-1$. In a real material, however, maximum screening is achieved at $\lambda(T=0)$. \textit{This is the key consideration for extracting the London penetration depth from magnetic susceptibility in ultrathin films.} For a bulk crystal, and even regular thin films ($d\sim100$~nm), the resulting deficit, $(1-N)\chi(T\to0)\simeq -1+\lambda(0)/R$, is of order $10^{-2}$ and is legitimately ignored. In our ultrathin film, it is $30\%$, and therefore cannot be neglected.

Figure~\ref{fig:chi(lambda)} shows the normalized $\chi(\lambda)$ of the film, obtained by a finite-element solution of the London equation (COMSOL) for a disc of radius $a$ and thickness $d$. For $d\ll\lambda$ the problem reduces to that of a zero-thickness sheet, in which $\chi/\chi_0$ depends on $\lambda$ only through $\Lambda/a$, with Pearl's length $\Lambda=2\lambda^2/d$. This sheet problem was solved for strips, discs and rectangles by Brandt, as the Meissner-state limit of the linear ac response~\cite{brandt1994,brandt1994a,brandt1995}; for the disc his tabulated pole expansion~\cite{brandt1994} and our finite-element curve agree to better than
$0.1\%$. Two-coil mutual inductance rests on the same electrodynamics in a different geometry: a film much larger than the coils is driven from one side, the coil--coil mutual inductance $M(\Lambda)$ is computed for the particular coils and inverted, and the film edges are kept away from the coils rather than modeled~\cite{fiory1988,turneaure1996,turneaure1998,paget1999}, except in recent work on small films~\cite{zhang2021}. By contrast, the
response of the whole sample (our case here) is dominated by the edges. Closest to the present work, Chen \textit{et al.} computed the uniform-field susceptibility of a square film as a function of $\Lambda$ and inverted \textit{ac} susceptibility data of a $250$~nm YBa$_2$Cu$_3$O$_{7-\delta}$ film to obtain $\lambda(T)$~\cite{chen2014}.

Our analysis differs in one essential respect. A tunnel-diode resonator measures a frequency shift proportional to $\Delta\chi$, with a calibration constant that must be supplied externally. For bulk crystals it is fixed by assuming that the signal at base temperature equals the ideal ($\lambda\to0$) response $\chi_0$, which errs by $\sim\lambda(0)/R$, typically $10^{-3}$ or less. For the present films this assumption fails: with $\lambda(0)\approx1.14~\mu$m the Pearl length at base temperature is a substantial fraction of the sample size, and $\chi(\lambda_0)/\chi_0\approx0.7$. We therefore fix the calibration constant with the computed $\chi(\lambda_0)/\chi_0$ rather than with $\chi_0$; this is the step this work adds. A detailed treatment of the ultrathin geometry, the finite-thickness corrections and the
dimensional crossovers will be published separately.

For reference, the full calibration procedure to extract the London penetration depth from measured linear \textit{global} magnetic susceptibility of the \textit{whole sample} of a thin film in a perpendicular magnetic field is then:\\
\begin{enumerate}
\item Compute $\chi(\lambda)$ for the given geometry; this is material-independent, and by the $\Lambda$-scaling above it need be done only once as a function of $\Lambda/a$. (If a full calculation is not available, an excellent approximate solution for a 2D sheet-current $\chi(\lambda)$ for arbitrary $\Lambda$ is given by Brandt in Ref.~\cite{brandt1994}).
\item Obtain $\lambda(0)$ for the material under study, from experiment or from theory.
\item Read the normalization constant $(1-N)\chi(T \to 0,\lambda(0))$ off the curve of step (1). \emph{This step replaces the conventional value $-1$ and is what makes the inversion applicable to a relative-signal probe.}
\item Normalize the measured susceptibility $\chi(T)$ to this value.
\item Invert numerically the calculated $\chi (\lambda) \to \lambda (\chi)$.
\item Apply the inversion to the normalized $\chi(T)$ to obtain $\lambda(T)$.
\end{enumerate}

For the present samples we take $\lambda(0)$ from the Drude--London estimate of Table~\ref{tab:lambda}. With $\lambda(0)\approx1.14\,\mu\mathrm{m}$ for FeTe one has $\Lambda=\lambda^{2}/d=0.13$~mm, hence $\Lambda/a=0.051$ and a normalization constant of $-0.70$, as marked in Fig.~\ref{fig:chi(lambda)}. Two sources of uncertainty deserve comment. The first is the thickness: since $\chi$ depends on $\lambda$ and $d$ only through Pearl's $\Lambda$, the range $d=6$--$10$~nm is \emph{exactly} equivalent to holding $d=10$~nm and rescaling $\lambda(0)$ to $0.88$--$1.14\,\mu\mathrm{m}$, which is the range explored below for the analysis of the superconducting gap; the thickness uncertainty therefore does not propagate into the conclusions at all. The second is $\lambda(0)$ itself. The interfacial layer is certainly depleted of carriers relative to bulk crystals, so the Drude--London value is an upper bound of carrier concentration, hence a lower bound of $\lambda(0)$; larger $\lambda(0)$ gives a smaller normalization constant, $|\chi(T\to0)|<0.7$. We repeated the entire analysis over $-0.98\le\chi(T\to0)\le-0.45$ (i.e.\ $0.19\,\mu\mathrm{m}\le\lambda(0)\le2.02\,\mu\mathrm{m}$) in Sec.~\ref{sec:gap}, and found that the conclusions are unchanged.

\begin{figure}[tb]
\includegraphics[width=8cm]{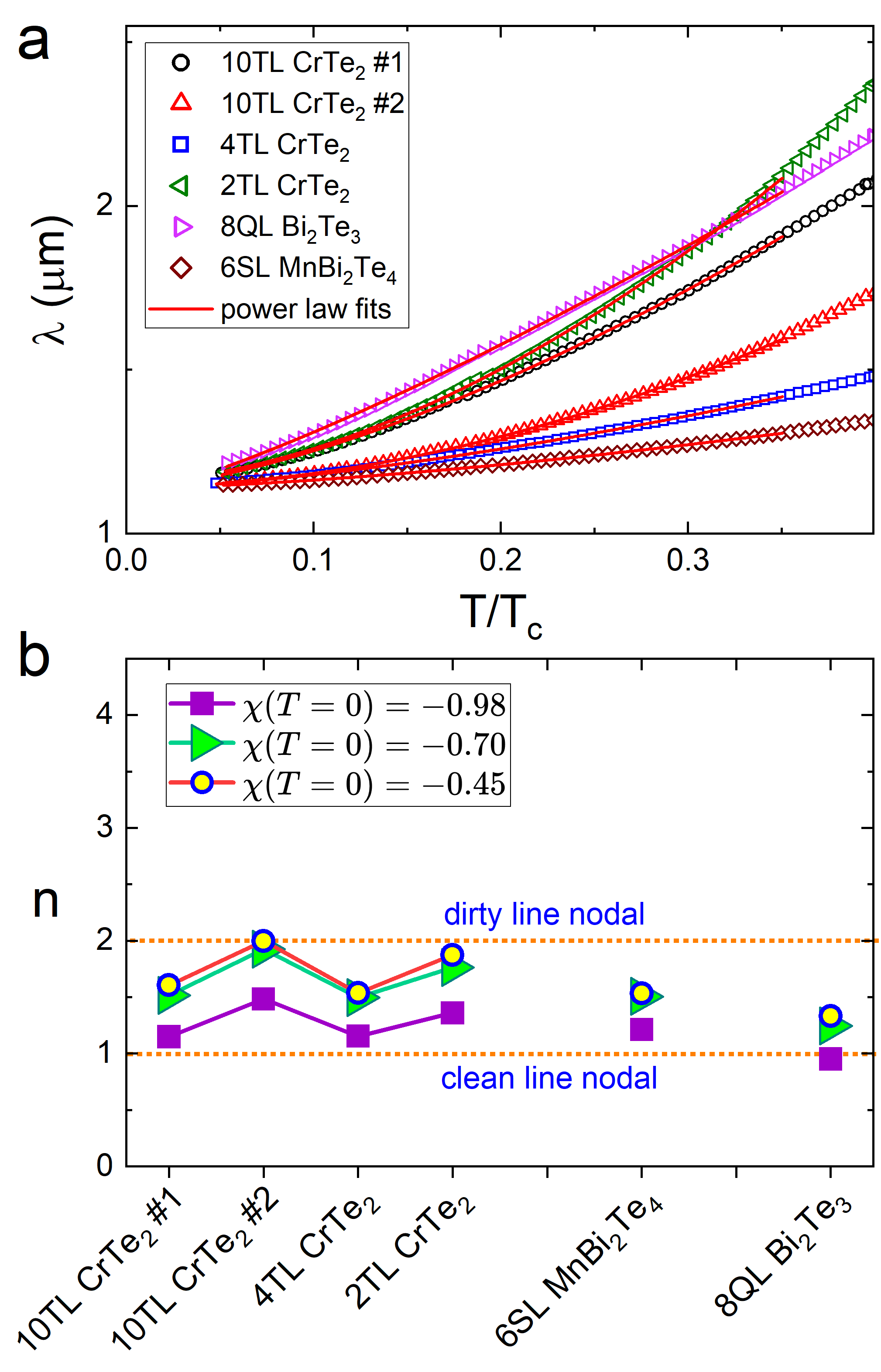}
\caption{ (a) Low-temperature $\lambda (T/T_c)$ analyzed using power-law behavior, $\Delta \lambda (t) \propto t^n$, which in recent years has become a standard practice to distinguish nodal and nodeless order parameters \cite{prozorov2011}. Fits are over $0.05 \le T/T_c \le 0.35$. (b) The obtained exponents $n$ for the studied materials. Different symbols correspond to different $\lambda(T=0)$ shown in the legend. The values between 1 and 2 are consistent with a line nodal order parameter with various degrees of non-magnetic disorder scattering, though a strongly anisotropic nodeless gap is not excluded.
}
\label{fig:lambda(t)_and_n}
\end{figure}

The normalized magnetic susceptibilities, $\chi(T)/\chi(T\to0)$
for the six heterostructures are shown in Fig.~\ref{fig:chi(t)_and_lambda(t)}(a), where $\chi(T\to0)=-0.7$ was estimated as shown in Fig.~\ref{fig:chi(lambda)} for $\lambda(0)=1.14\,\mu\mathrm{m}$. The calibrated London penetration depths are shown in Figure \ref{fig:chi(t)_and_lambda(t)}(b).  Note how $\lambda (T)$ curves increase sharply near $T_c$ as expected for intrinsic $\lambda (T)$ (of course this procedure is only applicable some distance from $T_c$). However, there are differences in the behavior between different samples. For example, the broader $\lambda (T)$ for the 2TL$-$CrTe$_2$ sample correlates with its lower $T_c$ ($\approx 9.5$~K versus $11-12$~K for the others).

\begin{figure*}[tb]
\includegraphics[width=17cm]{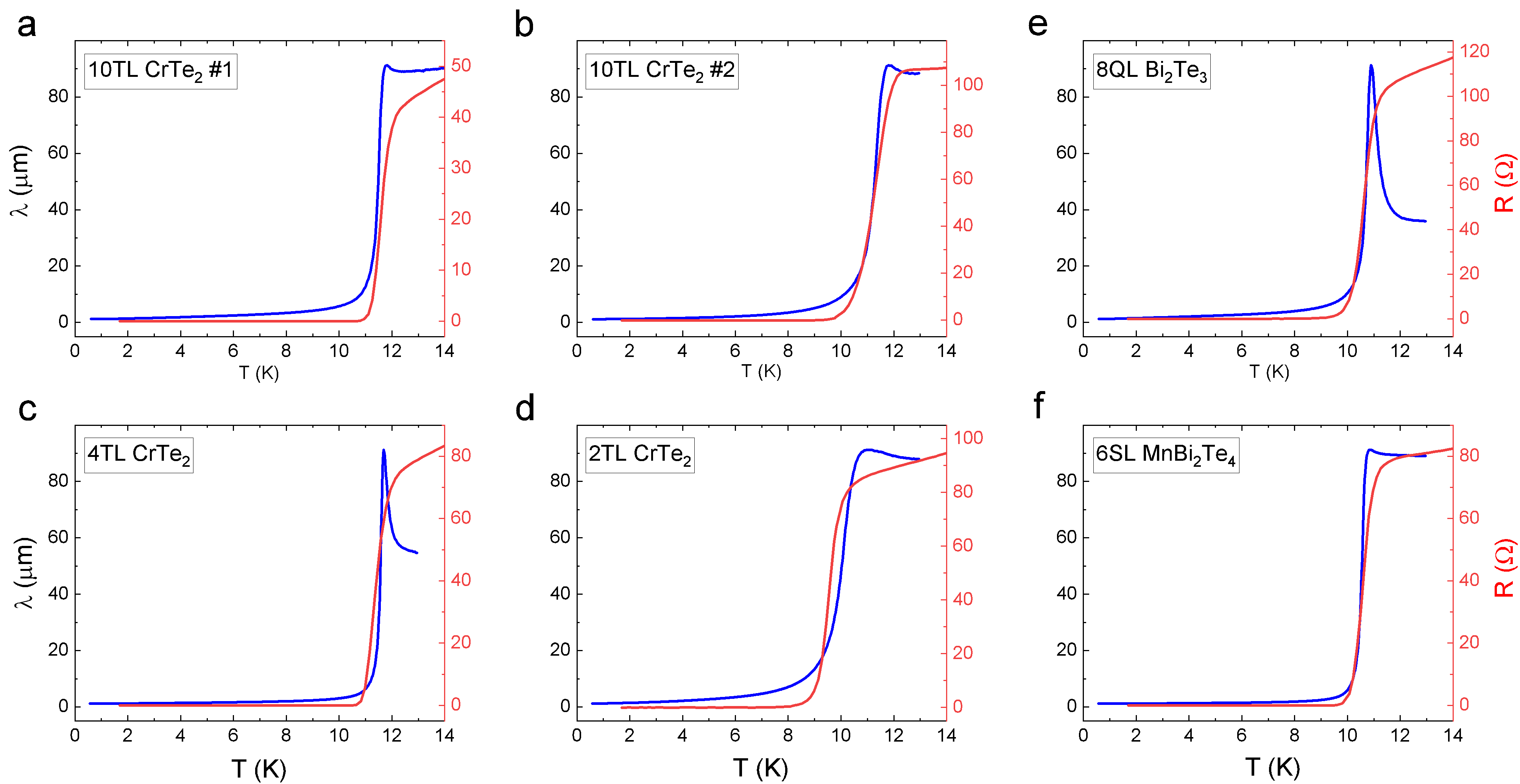}
\caption{The converted $\lambda (T)$ curves are plotted together with $R(T)$ curves, left and right Y-axes, respectively. The transition widths in both measured quantities practically coincide in all samples, except for the 2TL-$\textrm{CrTe}_2$ sample with the thinnest top layer.
}
\label{fig:lambda(t)_and_R(t)}
\end{figure*}

Next, we compare the superconducting transitions from $\lambda (T)$ and from the measured resistance shown in Fig.~\ref{fig:lambda(t)_and_R(t)}, and find excellent agreement between them not only in the onsets, but also in the sharpness of the transition curves, except for the 2TL-CrTe$_2$ sample. While the magnetic susceptibility shown in Fig.~\ref{fig:df(T)_and_R(T)} is much broader than $R(T)$, which is conventionally blamed on inhomogeneity of the superconducting state, here it is clear that this is just the expected effect of ultrathin geometry, and the extracted $\lambda(T)$ follows $R(T)$ very well.

This confirms that superconductivity in the studied heterostructures is a real macroscopic state of several interfacial layers across the whole sample, and the broad transition of the measured $\chi (T)$ does not necessarily signify inhomogeneity and is expected in the ultrathin limit with a large London penetration depth.

\section{Possible Superconducting Gap Structure}
\label{sec:gap}

Next, we use the extracted $\lambda (T)$ to examine the superconducting gap structure. To begin, we employ the now commonly used fit of the low-temperature variation to a power law, $\Delta\lambda = \lambda(T)-\lambda(0) = A\,t^{n}$ with $t=T/T_c$, as shown in Figure \ref{fig:lambda(t)_and_n}(a). The resulting exponents, $n$, are shown in Figure \ref{fig:lambda(t)_and_n}(b). The exponent $n$ varies between about 1 and 2, regardless of whether the top-layer thickness decreases or a different top-layer material is used. Because $\lambda(T=0)$ determines the calibration constant, its uncertainty can in principle affect both the extracted $\lambda(T)$ and the fitted exponent $n$. We explore this possibility by extracting $\lambda(T)$ from the calibration obtained with different $\lambda(T=0)$. Since this value determines the $\chi(T \to 0, \lambda(0))$ extrapolation, we repeated the analysis over a broad range of these values, from $\chi(0)=-0.98$ to $-0.45$, corresponding to $\lambda(0)= 0.19\,\mu\mathrm{m}$ and $2.02\,\mu\mathrm{m}$, respectively. Even with such a considerable range, the exponents $n$ remain between about 1 and 2.

For a fully gapped superconductor, an exponentially activated $\Delta\lambda(T)$ is practically indistinguishable from a power law with $n \gtrsim 3$, and fits to real data on nodeless superconductors typically return $n > 3$. A line-nodal superconductor exhibits $n=1$ in the clean case and approaches $n=2$ in the dirty limit \cite{hirschfeld1993}, although how much disorder is needed depends decisively on the scattering phase shift, as shown below. The intermediate values of $1\leq n \leq 2$ are expected from resonant disorder scattering, but can also be due to a strongly anisotropic nodeless superconducting gap or a complicated multi-gap superconducting state when the minimum temperature of the experiment is above the gap minimum (in temperature units) \cite{cho2022,teknowijoyo2016}.

Since the exponents fall between the clean and dirty nodal limits, it is useful to have a quantitative measure of where within that range they lie. Two conventions are needed first. We quantify disorder by the dimensionless scattering rate $\Gamma=\hbar/(2\pi k_BT_{c0}\tau)$ and write $t_c=T_c/T_{c0}$ for the reduced transition temperature, where $\tau$ is the quasiparticle scattering time and $T_{c0}$ is the transition temperature the same material would have in the absence of pair-breaking disorder; normalizing by $T_{c0}$ rather than by $T_c$ makes $\Gamma$ depend on $\tau$ alone. Second, the strength of an individual scattering event is set by its phase shift, and we refer below to the two standard extremes, the weak-scattering (Born) and the resonant (unitarity) limits, which differ qualitatively in the residual quasiparticle density of states they generate at a line node.

The Hirschfeld--Goldenfeld interpolation formula~\cite{hirschfeld1993}, derived for resonant scatterers and written in the reduced temperature $t=T/T_c$ used throughout, reads $\Delta\lambda=a\,t^{2}/(t+t^{\ast})$, where $t^{\ast}=T^{\ast}/T_c$ is the impurity crossover scale below which $\Delta\lambda$ becomes quadratic. Fitting the power law $\Delta\lambda=A\,t^{n}$ to this form over our window $0.05\le t\le0.35$ gives, to better than $10\%$ for $1.2\lesssim n\lesssim1.9$,
\begin{equation}
t^{\ast}\simeq 0.215\,\frac{n-1}{2-n},
\label{eq:Tstar}
\end{equation}
\noindent so that $n=1.2$, $1.5$ and $1.8$ correspond to $t^{\ast}=0.054$, $0.215$ and $0.86$; the coefficient is set by the fitting window and should be quoted together with $t^{\ast}$. Our exponents therefore place $T^{\ast}$ between roughly $0.05\,T_c$ and $T_c$.

An absolute pair-breaking rate cannot be extracted from this. Although our $T_c$ lie below the $13.5$~K of stoichiometric bulk $\mathrm{FeTe}$~\cite{yan2026stoichiometric}, in the ultrathin limit $T_c$ is suppressed by a reduced density of states as much as by scattering, and the same reduction raises $\lambda(0)$, since $T_c\propto\exp[-1/N(0)V]$ while $\lambda^{-2}(0)$ is proportional to the carrier density; the clean-limit $T_{c0}$ of the interfacial layer is therefore unknown, and with it $\Gamma$ itself. A bound survives that needs neither $T_{c0}$ nor $\tau$. In the Hirschfeld--Goldenfeld theory, the crossover temperature is fixed by the scattering-rate parameter and the gap maximum, $k_BT^{\ast}\simeq0.83\,(\Gamma_E\Delta_0)^{1/2}$ with $\Gamma_E=\hbar/2\tau=\pi k_BT_{c0}\Gamma$~\cite{hirschfeld1993}, and the same $\Gamma_E$ is the Abrikosov--Gor'kov pair-breaking parameter, which destroys a line-nodal state at $\Gamma_c=0.882/\pi\simeq0.28$ in the present units, the value $0.281$ found numerically in Ref.~\onlinecite{kogan2013} in this same convention. Taking $\Delta_0=2.14\,k_BT_c$ for a weak-coupling $d$-wave gap, both $\tau$ and $T_{c0}$ drop out of the ratio, leaving
\begin{equation}
\frac{\Gamma}{\Gamma_c}=\frac{t_c\,t^{\ast2}}{1.30}\;\le\;\frac{t^{\ast2}}{1.30}\;\simeq\;0.036\left(\frac{n-1}{2-n}\right)^{\!2},
\label{eq:bound}
\end{equation}
\noindent where the inequality follows from $t_c\le1$, that is, from the fact that disorder can only lower $T_c$. This gives $\Gamma/\Gamma_c\lesssim0.08$ for $n\lesssim1.6$, covering most of our samples, so these films are far from the disorder that would destroy a nodal gap. Equation~(\ref{eq:bound}) presumes the dilute limit, $k_BT^{\ast}\ll\Delta_0$, and is quantitative only for $n\lesssim1.7$. Equations~(\ref{eq:Tstar}) and~(\ref{eq:bound}) presume resonant scattering, and the alternative is not a minor caveat, because the two limits behave oppositely. Kogan \emph{et al.}~\cite{kogan2013} solved the Born problem for a $d$-wave order parameter using the same reduced variables adopted here, and found that transport and spin-flip rates enter only through their sum, that $\rho(T/T_c)$ stays close to its clean-limit form --- slope $4/3$ at $T_c$ --- for $\Gamma$ up to about half of $\Gamma_c$, and that the linear low-temperature $\lambda(T)$ becomes quadratic only once $T_c$ is suppressed below $T_{c0}/3$. In the unitary limit the same calculation reproduces the Hirschfeld--Goldenfeld behaviour, in which even weak scattering removes the linear signature.

This sharpens the argument rather than weakening it. Born scattering cannot account for our exponents: $t_c\le1/3$ would require $T_{c0}\gtrsim30$~K for the $T_c$ measured here, more than twice the $13.5$~K of stoichiometric bulk $\mathrm{FeTe}$~\cite{yan2026stoichiometric} and the highest yet reported in this family, while the reduced density of states of an ultrathin layer moves $T_{c0}$ down rather than up. It also corresponds to $\Gamma/\Gamma_c\approx0.8$, an order of magnitude above the bound of Eq.~(\ref{eq:bound}). Either the scatterers are resonant, in which case Eq.~(\ref{eq:bound}) applies and these are weakly disordered nodal superconductors, or the exponents are not a disorder effect at all and reflect gap anisotropy, multi-band structure, or the finite $0.5$~K floor of the experiment. Separating a nodal gap from a deeply anisotropic nodeless one therefore requires an independent measure of the strength of the scattering potential, which the present data do not provide.

An entirely independent recent local measurement on stoichiometric $\mathrm{FeTe}$ using scanning SQUID reports a power-law exponent of $\approx 1-1.5$ down to $0.02\,T_c$ \cite{li2026nodal}, consistent with the range found here and with the emerging conclusion that the interfacial superconductivity originates in $\mathrm{FeTe}$ forming at the interface.

\begin{figure*}[tb]
\includegraphics[width=17cm]{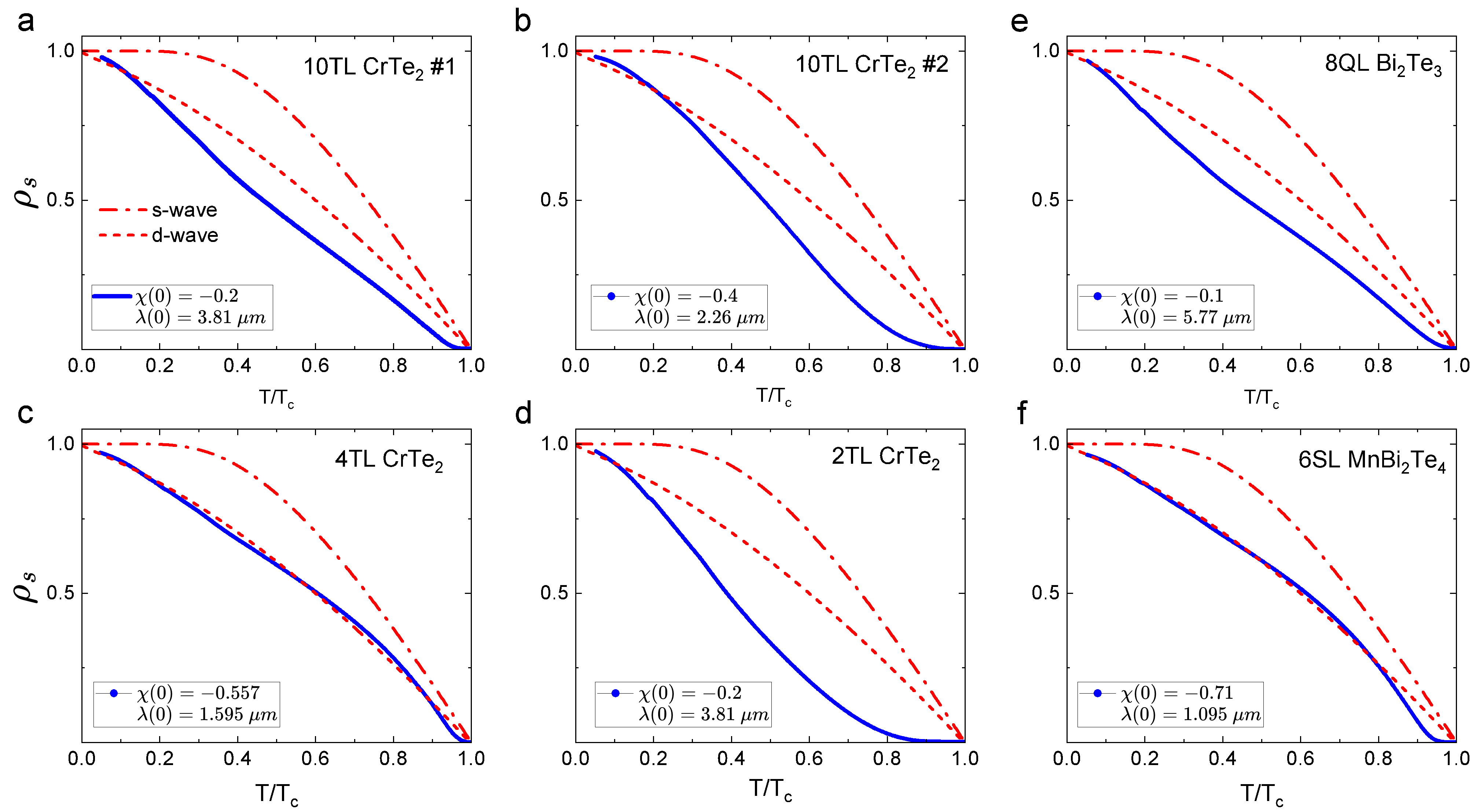}
\caption{(a)-(f) Superfluid density versus normalized temperature for the six systems. Blue solid lines are the data; the red dash-dotted and dashed curves are the clean-limit isotropic $s$-wave and nodal $d$-wave calculations. The $\rho_s$ for 4TL $\mathrm{CrTe}_2$ and 6SL $\mathrm{MnBi}_2\mathrm{Te}_4$ fit well with the nodal $d$-wave curve. Extreme values of $\lambda(T=0)$ do not yield a good fit for the rest of the films. Unlike Figure \ref{fig:lambda(t)_and_n}, which uses a single normalization for all six samples, here $\lambda(0)$ is a per-sample free parameter; the value used is given in each panel.
}
\label{fig:rho_s(t)}
\end{figure*}

To clarify this further, we note that the full temperature range behavior of the superfluid phase stiffness (superfluid density) is dramatically different between fully gapped and nodal cases.
Blue solid lines in Figure \ref{fig:rho_s(t)} show the superfluid density, $\rho_s(t)=(\lambda(0)/\lambda(t))^2$, obtained from the measured $\lambda(T)$ in the studied $\mathrm{FeTe}$ heterostructures; the red curves are the clean-limit isotropic $s$-wave and nodal $d$-wave calculations. Since $\lambda(0)$ is not known independently, we treat it as a single free parameter per sample and report the value required to bring $\rho_s(t)$ onto each candidate curve; the resulting $\lambda(0)$ can then be compared with the independent Drude--London estimate of Table~\ref{tab:lambda} as a consistency test. We find that the $\rho_s$ for 4TL $\mathrm{CrTe}_2$ and 6SL $\mathrm{MnBi}_2\mathrm{Te}_4$ are in excellent agreement with the clean-limit $d$-wave order parameter --- an identification that is robust to moderate disorder, since $\rho(T/T_c)$ is nearly universal for Born scattering up to $\Gamma\approx\Gamma_c/2$~\cite{kogan2013} --- and that this requires $\lambda(T=0)=1.60$ and $1.10\,\mu\mathrm{m}$, respectively --- the same order as our reference value of $1.14\,\mu\mathrm{m}$, see Table~\ref{tab:lambda}. The other systems are likely more complicated due to a mix of several factors, such as spatial inhomogeneity, which cannot be excluded in these heterostructures \cite{tkac2024,li2026nodal}. A Berezinskii--Kosterlitz--Thouless (BKT) transition is also frequently invoked in this context, and it is worth checking quantitatively whether it can play that role here; as shown in Sec.~\ref{sec:bkt}, it cannot, but the same estimate turns out to constrain $\lambda(0)$ usefully.

\section{Two-dimensional phase stiffness and the BKT scale}
\label{sec:bkt}

A film this thin is a two-dimensional superconductor in the strict sense, so its phase stiffness is finite and vortex--antivortex unbinding must be considered. The areal stiffness is
\begin{equation}
J=\frac{\hbar^{2}d}{4\mu_0e^{2}\lambda^{2}},\qquad k_BT_{\rm BKT}=\frac{\pi}{2}J(T_{\rm BKT}),
\label{eq:bkt}
\end{equation}
the second relation being the Nelson--Kosterlitz universal-jump condition~\cite{nelson1977}. Evaluating $J$ with the zero-temperature penetration depth gives a bare scale $T^{0}_{\rm BKT}=\pi\hbar^{2}d/(8\mu_0e^{2}k_B\lambda^{2}(0))$; solving Eq.~(\ref{eq:bkt}) self-consistently with the clean $d$-wave $\rho_s(t)$ of Fig.~\ref{fig:rho_s(t)} gives the actual transition. Both are listed in Table~\ref{tab:bkt} for the $\lambda(0)$ values used in Fig.~\ref{fig:rho_s(t)}, at $d=10$~nm and $T_c=11$~K.
\begin{table}[htb]
\centering
\begin{tabular}{cccccl}
\hline
$\lambda(0)$ & $\Lambda/a$ & $\chi(T\!\to\!0)$ & $T^{0}_{\rm BKT}$ & $T_{\rm BKT}/T_c$ & sample \\
($\mu$m) & & & (K) & & \\
\hline
0.19 & 0.0014 & $-0.98$ & 2700 & 1.00 & scan limit, Fig.~\ref{fig:lambda(t)_and_n}(b) \\
1.095 & 0.047 & $-0.71$ & 82 & 0.91 & 6SL MnBi$_2$Te$_4$ \\
1.14 & 0.051 & $-0.70$ & 76 & 0.90 & Table~\ref{tab:lambda} reference \\
1.595 & 0.100 & $-0.56$ & 39 & 0.82 & 4TL CrTe$_2$ \\
2.26 & 0.201 & $-0.40$ & 19 & 0.69 & 10TL CrTe$_2$ \#2 \\
3.81 & 0.570 & $-0.20$ & 6.8 & 0.42 & 10TL \#1, 2TL CrTe$_2$ \\
5.77 & 1.307 & $-0.10$ & 2.9 & 0.23 & 8QL Bi$_2$Te$_3$ \\
\hline
\end{tabular}
\caption{Two-dimensional phase stiffness for $d=10$~nm. $T^{0}_{\rm BKT}$ is the bare Nelson--Kosterlitz scale evaluated with $\lambda(0)$; $T_{\rm BKT}/T_c$ is the self-consistent solution of Eq.~(\ref{eq:bkt}) using a clean $d$-wave $\rho_s(t)$ and $T_c=11$~K. $\Lambda=\lambda^{2}(0)/d$ and $a=2.55$~mm.}
\label{tab:bkt}
\end{table}

Two conclusions follow. First, for the physically sensible values $\lambda(0)\approx1.1$--$1.6\,\mu\mathrm{m}$ the bare scale is $4$--$7$ times $T_c$, so the transition is not stiffness-limited: BKT narrows the last $10$--$20\%$ of the transition and does nothing at all below that. It therefore cannot be the reason $\rho_s(t)$ in Fig.~\ref{fig:rho_s(t)}(a), (b), (d) and (e) falls below the $d$-wave curve over the entire temperature range. Second, and more usefully, the argument can be inverted. The diamagnetic and resistive onsets coincide in five of the six samples (Fig.~\ref{fig:lambda(t)_and_R(t)}), which requires $T_{\rm BKT}\gtrsim0.9\,T_c$ and hence, from Table~\ref{tab:bkt}, $\lambda(0)\lesssim1.2\,\mu\mathrm{m}$ at $d=10$~nm (the bound scales as $\sqrt{d}$, giving $0.9\,\mu\mathrm{m}$ at $d=6$~nm). The much larger $\lambda(0)$ needed to force the remaining samples onto a gap curve are excluded on phase-stiffness grounds alone: a film with $\lambda(0)=5.8\,\mu\mathrm{m}$ and $d=10$~nm would lose phase coherence near $2.5$~K and could not screen at $11$~K. This bound is entirely independent of the calibration of Sec.~\ref{sec:calib}, and it lands on the Drude--London estimate of Table~\ref{tab:lambda}.

Both statements can be tested against an existing measurement. He \emph{et al.} identify a BKT transition in $\mathrm{Bi}_2\mathrm{Te}_3(7\,\mathrm{QL})/\mathrm{FeTe}$ from current--voltage characteristics and a Halperin--Nelson fit to $R(T)$, obtaining $T_{\rm BKT}=10.1$~K against a mean-field $T_c=11.1$~K, together with a superconducting thickness $d_{\rm sc}=7.0\pm1.1$~nm from the upper critical fields~\cite{he2014}. Solving Eq.~(\ref{eq:bkt}) for that $T_{\rm BKT}/T_c=0.91$ at their own $d_{\rm sc}$ gives $\lambda(0)=0.90\,\mu\mathrm{m}$, with the range $0.83$--$0.97\,\mu\mathrm{m}$ spanned by their thickness uncertainty. This is within $20\%$ of the Drude--London estimate of Table~\ref{tab:lambda}, close agreement given that the latter uses bulk carrier density and effective mass, and it is smaller by a factor of six than the $5.8\,\mu\mathrm{m}$ that Fig.~\ref{fig:rho_s(t)}(e) would require for our own $\mathrm{Bi}_2\mathrm{Te}_3$ sample. An independently measured BKT transition in the same material system therefore supports both the scale of $\lambda(0)$ and the exclusion of the large values.

Two points should be stated plainly. We do not claim to observe a BKT transition here: $\rho_s(t)$ in Fig.~\ref{fig:rho_s(t)} shows no universal jump, and we have no current--voltage characteristics with which to look for one. The estimate is used only to bound $\lambda(0)$, and for that purpose it is conservative --- evaluating Eq.~(\ref{eq:bkt}) with the bare stiffness neglects the vortex fugacity, which depresses $T_{\rm BKT}$ further and therefore tightens the bound rather than relaxing it. It is also insensitive to pairing symmetry, since the universal-jump condition is a property of the phase field alone and the gap enters only through the near-$T_c$ slope of $\rho_s$.


\section{Discussion}

Our main result is that $f$TDR measurements establish the macroscopic, whole-sample Meissner--London diamagnetic response of six $\mathrm{FeTe}$-based heterostructures. In all of them $\chi(T)$ is far broader than $R(T)$, a signature usually attributed to inhomogeneity and currently the prevailing interpretation in ultrathin nickelates~\cite{harvey2025} and cuprates~\cite{shi2024}, and visible largely unexplained in other techniques~\cite{mandal2020,minhaj1994,kamlapure2010}. We show instead that most of it follows from the extreme ultrathin geometry combined with the large London penetration depth typical of low carrier density, and we give a quantitative procedure that removes it and returns $\lambda(T)$.

This does not exclude inhomogeneity. Cross-sectional TEM of these heterostructures shows clean epitaxial growth with atomically sharp interfaces~\cite{yan2025meissner,yuan2024coexistence,yi2023dirac}, but compositional, strain or stoichiometry variations at the interface could still produce it, and scanning-SQUID imaging of stoichiometric FeTe does resolve micrometer-scale variation in both stiffness and $T_c$~\cite{li2026nodal}, with multiphase superconductivity imaged at the FeTe/Bi$_2$Te$_3$ interface~\cite{tkac2024}. Such long-wavelength variation need not smear the transition, however, and we show that the broader $\chi(T)$ does not \emph{require} inhomogeneity, and the global $\lambda(T)$ --- a weighted average over the whole sample --- retains a well-defined power-law form with similar exponents in all six samples.

The results therefore point to intrinsic properties of the interfacial $\mathrm{FeTe}$ layer, which recent work shows to superconduct once excess iron is removed~\cite{yan2026stoichiometric}. Our sample set also bears on the original motivation for these heterostructures: the three overlayers span three of the four combinations of band topology and magnetism --- $1T-\mathrm{CrTe}_2$ ferromagnetic and topologically trivial~\cite{freitas2015}, $\mathrm{Bi}_2\mathrm{Te}_3$ a non-magnetic topological insulator~\cite{chen2009}, $\mathrm{MnBi}_2\mathrm{Te}_4$ both~\cite{Li2019,Zhang2019,Otrokov2019} --- yet $T_c$, the exponent $n$ and the bound on $\lambda(0)$ are indistinguishable across all six, so neither the topology nor the magnetism of the overlayer leaves a measurable imprint. Together with the evidence that interfacial Te stoichiometry controls the effect~\cite{yao2025mystery}, this suggests the overlayer acts chemically rather than electronically --- most plausibly as a Te reservoir suppressing interstitial Fe in the topmost layers. Consistently, the lowest $T_c$ and the only onset mismatch occur for the thinnest overlayer (2TL $\mathrm{CrTe}_2$), suggesting that $T_c$ tracks overlayer thickness rather than identity --- though on a single sample.

This relocates the topological question rather than closing it. A global magnetic probe is blind to a helical surface state by construction: its counterpropagating, spin-momentum-locked channels carry compensating charge currents and contribute nothing essentially to the screening we measure. What the present data exclude is a mechanism in which the overlayer's topology produces the pairing; they say nothing about what that pairing may induce in an adjacent surface state. That question is a spectroscopic one --- the signature sought in a proximitized topological surface state is a zero-bias peak from a Majorana mode \emph{inside} the induced gap, rather than the gap itself~\cite{fu2008} --- and answering it requires local probes on the overlayer surface~\cite{li2026nodal}. It is also a different problem from the one Fu and Kane considered, since the parent superconductor here is not a conventional $s$-wave one. A decisive test of the chemical picture, by contrast, is straightforward and has not been done: capping with a chemically inert, Te-free layer.

On the gap itself, the strongest statement is not the low-temperature exponent but the full-range superfluid density: $\rho_s(T)$ for two of the six samples follows the standard $d$-wave curve over the entire temperature range, with the single adjustable parameter $\lambda(0)$ taking the $1.1$--$1.6\,\mu$m independently expected from Table~\ref{tab:lambda}. Such values are unremarkable for a low-carrier-density, mass-renormalized system like $\mathrm{FeTe}$~\cite{lin2025,tamai2010}, and are consistent with the reduced density of states that also lowers $T_c$ in the ultrathin limit (Sec.~\ref{sec:gap}): the two observations share one cause rather than being independent. Scanning tunneling spectroscopy on these heterostructures reports predominantly V-shaped spectra with broad coherence peaks~\cite{qin2020,manna2017,yi2024,yuan2024coexistence,li2026nodal}, and twofold-symmetric superconductivity has been seen in Bi$_2$Te$_3$/FeTe$_{0.55}$Se$_{0.45}$~\cite{chen2018twofold}; none fixes the gap structure, but all point away from an isotropic gap. The remaining four samples would require $\lambda(0)$ that Sec.~\ref{sec:bkt} excludes on phase-stiffness grounds, so their near-linear $\rho_s(t)$ is most likely a distribution of $T_c$ within the sample rather than a gap function at all.

In summary, we measured the global Meissner response of interfacial superconductivity in three types of $\mathrm{FeTe}$ heterostructure using a frequency-domain tunnel-diode resonator, and developed a calibration that recovers the intrinsic $\lambda(T)$ from a susceptibility broadened by the ultrathin geometry. The extracted superfluid density is inconsistent with a conventional isotropic $s$-wave gap and points to a strongly anisotropic order parameter with line nodes or deep minima, in a condensate that is spatially coherent across the whole sample.

\section{Methods}
\subsection{Molecular-beam epitaxy growth}
All FeTe-based heterostructures used in this work are grown on $3\,\mathrm{mm} \times 10\,\mathrm{mm}$, 0.5~mm-thick insulating SrTiO$_3$(100) substrates in a commercial MBE chamber (Scienta Omicron Lab10) with a base pressure below $\sim 2 \times 10^{-10}$~mbar. The SrTiO$_3$(100) substrates are first soaked in $\sim 80\,^\circ$C deionized water for $\sim 2$~h and subsequently in a diluted hydrochloric acid solution ($\sim 4.5\%$ w/w) for $\sim 2$~h. These SrTiO$_3$(100) substrates are then annealed in a tube furnace under flowing high-purity oxygen gas at $\sim 974\,^\circ$C for $\sim 3$~h. This heat treatment produces a suitable SrTiO$_3$(100) surface for the MBE growth of FeTe-based heterostructures. The heat-treated SrTiO$_3$(100) substrates are loaded into the MBE chamber and outgassed at $\sim 600\,^\circ$C for $\sim 1$~h before MBE growth. High-purity Fe (99.995\%), Te (99.9999\%), Cr (99.999\%), Bi (99.9999\%), and Mn (99.9998\%) are evaporated from Knudsen effusion cells. The FeTe, 1T-CrTe$_2$, Bi$_2$Te$_3$, MnBi$_2$Te$_4$ layers are grown at substrate temperature of $\sim 340\,^\circ$C, $\sim 300\,^\circ$C, $\sim 210\,^\circ$C, and $\sim 270\,^\circ$C, respectively. The corresponding growth rates are $\sim 0.3$~UC/min for FeTe, $\sim 0.25$~TL/min for 1T-CrTe$_2$, $\sim 0.2$~QL/min for Bi$_2$Te$_3$, and  $\sim 0.2$~SL/min for MnBi$_2$Te$_4$. No additional protective capping layer is deposited. More details about the MBE growth and structural characterization are provided in Refs.~\cite{yan2025meissner, yi2023dirac, yuan2024coexistence}.

After MBE growth, each heterostructure is divided into two companion pieces, each with an effective film area of $\sim 3\,\mathrm{mm} \times 4\,\mathrm{mm}$. One piece is used for the $f$TDR measurements performed at Iowa State, while the other is used for the electrical transport measurements performed at Penn State.
\subsection{Planar thin-film $f$TDR and transport measurements}
The self-oscillating tunnel diode resonator ($f$TDR) consists of two planar coils in a Helmholtz configuration, which act as the inductor, and it operates around 10~MHz. A thin-film sample of lateral dimensions $3\,\mathrm{mm} \times 4\,\mathrm{mm}$ is inserted between these coils, which are designed to accommodate the 0.5 mm substrate. The resonant frequency shift is measured by sweeping the temperature from $500\,\mathrm{mK}$ to $12\,\mathrm{K}$. The amplitude of the ac excitation field is $H_{ac} \approx$ 20 mOe, so the response is in the linear Meissner regime and no vortices are created. For a thin film in a perpendicular field the relevant threshold is not the bulk $H_{c1}$ but the field at which the edge sheet current reaches the depairing value; for our geometry this gives a first-entry field of order $0.2$~Oe, an order of magnitude above the excitation used. Examples of raw data are shown in Fig.~\ref{fig:df(T)_and_R(T)} and the calibration is described in the main text. More details about the $f$TDR setup and the measurement procedure are described elsewhere \cite{Van1975,prozorov2000a,prozorov2021,giannetta2022london}.

The $\sim 3\,\mathrm{mm} \times 4\,\mathrm{mm}$ MBE-grown FeTe-based heterostructure films are scratched into a Hall-bar geometry using a computer-controlled motorized probe station. The effective area of the Hall-bar device is approximately $\sim 1\,\mathrm{mm} \times 0.5\,\mathrm{mm}$. The electrical contacts are made by pressing indium spheres onto the Hall bar. Electrical transport measurements are conducted using a Physical Property Measurement System (PPMS, Quantum Design DynaCool, 1.7~K, 9~T). The excitation current is 1~$\mu$A for all $R$--$T$ measurements. To minimize oxidation, all samples are measured within 30~min after being taken out of the MBE chamber.
\begin{acknowledgments}
We thank Katja Nowack and Cequn Li for useful discussions.
This work was supported by the U.S. Department of Energy (DOE), Office of Science, Basic Energy Sciences, Materials Science and Engineering Division. Ames National Laboratory is operated for the U.S. DOE by Iowa State University under contract DE-AC02-07CH11358. The MBE growth of FeTe-based heterostructures is supported by the DOE grant (DE-SC0023113). The electrical transport measurements performed at Penn State are supported by the ONR grant (N000142412133). C.-Z.C. acknowledges the support from the Gordon and Betty Moore Foundation’s EPiQS Initiative (GBMF9063 to C.-Z. C). 
\end{acknowledgments}
%

%

\end{document}